\documentstyle[12pt]{article}

\newcommand{\beq}{\begin{equation}}
\newcommand{\eeq}{\end{equation}}
\newcommand{\beqa}{\begin{eqnarray}}
\newcommand{\eeqa}{\end{eqnarray}}

\newcommand{\ie}{{\it i.e.}\ }
\newcommand{\eg}{{\it e.g.}\ }

\newcommand{\etal}{{\em et al.}}

\newcommand{\order}{{\cal O}}
\newcommand{\into}{\rightarrow}
\newcommand{\lra}{\leftrightarrow}

\newcommand{\ccbar}{c\bar{c}}
\newcommand{\QQbar}{Q\bar{Q}}

\newcommand{\ubar}{\bar{u}}
\newcommand{\dbar}{\bar{d}}
\newcommand{\jp}{$J/\psi$}
\newcommand{\jps}{$J/\psi$\ }

\newcommand{\ppi}{p_\pi}
\newcommand{\ptr}{P_\perp}

\newcommand{\spt}{$P_\perp$}

\newcommand{\als}{\alpha_s}
\newcommand{\as}{$\alpha_s$}

\newcommand{\sla}[1]{{#1}\!\!\! / \,}
\newcommand{\state}{\; ^{2S+1}\! L_J}
\newcommand{\stot}{s_{\rm tot}}

\newcommand{\xf}{$x_F$}
\newcommand{\xfs}{$x_F$\ }

\newcommand{\ARNPS}[3]{\mbox{}Ann. Rev. Nucl. Part. Sci.
{\bf B{#1}}, {#2} ({#3})}
\newcommand{\NPB}[3]{\mbox{}Nucl. Phys. {\bf B{#1}}, {#2} ({#3})}
\newcommand{\PLB}[3]{\mbox{}Phys. Lett. {\bf B{#1}}, {#2} ({#3})}
\newcommand{\PR}[3]{\mbox{}Phys. Rev. {\bf {#1}}, {#2} ({#3})}
\newcommand{\PRL}[3]{\mbox{}Phys. Rev. Lett. {\bf {#1}}, {#2} ({#3})}
\newcommand{\PRD}[3]{\mbox{}Phys. Rev. {\bf D{#1}}, {#2} ({#3})}
\newcommand{\ZPC}[3]{\mbox{}Z. Phys. {\bf C{#1}}, {#2} ({#3})}
\newcommand{\ibid}[3]{\mbox{}{\em ibid.} {\bf {#1}}, {#2} ({#3})}

\begin{document}

\setlength{\baselineskip}{7mm}

\thispagestyle{empty}
\begin{flushright}
   \vbox{\baselineskip 16pt plus 1pt minus 1pt
         HU-TFT-95-16 \\
         March 1995 \\
         hep-ph/9503250 }
\end{flushright}

\renewcommand{\thefootnote}{\fnsymbol{footnote}}
\bigskip
\begin{center}
{\Large \bf Hadroproduction of Charmonium at Large \xf}

\vskip 1\baselineskip

M. V\"anttinen \\
{\normalsize \em Research Institute for Theoretical Physics } \\
{\normalsize \em P.O.Box 9, FIN-00014 University of Helsinki, Finland} \\
vanttinen@phcu.helsinki.fi

\vskip 1\baselineskip

\end{center}

\medskip

\renewcommand{\thefootnote}{\arabic{footnote}}
\setcounter{footnote}{0}

\begin{center}
{\bf ABSTRACT}
\end{center}

\vbox{\baselineskip 14pt \noindent We present a higher-twist
production mechanism for $\eta_c$ and $\chi_J$
charmonium at large momentum fraction \xfs in pion-nucleon
collisions. The higher-twist contribution is essentially
independent of \xfs and
therefore becomes dominant at large \xf, where
the leading-twist contribution falls off as
$(1-x_F)^3$.
We show that the higher-twist mechanism produces longitudinally
polarized $\chi_1$ and $\chi_2$. For the $\chi_2$, this is
clearly different from the leading-twist prediction of transverse
polarization. For the $\chi_1$, the polarization of the
leading- and higher-twist
contributions is qualitatively similar.}

\bigskip
\begin{center}
{\it Submitted to Physical Review D}
\end{center}
\newpage

\section{Introduction}

Heavy quark-antiquark bound states (quarkonium) were among the first
systems to be successfully studied using perturbative QCD.
Much of the quarkonium decay data was adequately explained by the
colour-singlet model \cite{Kwong,Kopke,Schuler}. In this model,
the quarkonium states are taken to be non-relativistic colour-singlet
$\QQbar$ states, and decay amplitudes are written as convolutions
of the quarkonium wavefunction with the lowest-order perturbative
amplitudes for $\QQbar$ annihilation.
It could then be expected that quarkonium states are dominantly produced
by the leading-twist mechanisms where the subprocess
(\eg $gg \into J/\psi + g$)
is related to a decay mechanism (\eg $J/\psi \into ggg$) by crossing
\cite{BergerJones,BaierZPC19}. However, the resulting predictions
are contradicted by several experiments, from fixed-target
\cite{Biino,Akerlof,AntoniazziPRL70} to
the highest collider energies \cite{CDFDaniels,CDFPapadimitriou}.

In this article, we shall concentrate on the production of charmonium
at low transverse momentum \spt\
and large longitudinal momentum fraction \xfs
in fixed-target pion-nucleon collisions.
We have earlier
pointed out that the anomalous dependence of the production cross section
on the mass number $A$ of a nuclear target \cite{Badier}
suggests the
importance of higher-twist, non-factorizable production mechanisms of the \jps
in this region \cite{HVS}.
In a recent work
\cite{VHBT}, we calculated the polarization of \jp's produced in pion-nucleon
reactions in the colour-singlet model at leading twist,
including contributions from direct
parton fusion into \jps and from radiative decays of
the $\chi_{1,2}$ charmonium states. We found a discrepancy between
these leading-twist predictions and experimental data \cite{Biino,Akerlof}.
The purpose of this paper is to present a specific higher-twist
mechanism for the production of $C=+1$ charmonia (the $\eta_c$ and $\chi_J$)
at large \xf.
We shall show that higher-twist mechanisms are likely to be dominant at large
\xfs and that their characteristic features
may explain some of the charmonium polarization data.

The polarization of \jps charmonium produced in pion-nucleon collisions
has been measured by the
Chicago-Iowa-Princeton collaboration \cite{Biino}
for $ 0.3 < x_F < 1.0$. They observe
an abrupt change from unpolarized production
to longitudinal polarization
at $x_F \approx 0.95$. This effect is reminiscent of the longitudinal
polarization effect \cite{Guanziroli,Conway} observed in the production of
dileptons in pion-nucleon collisions, which may be explained by a higher-twist
production mechanism first suggested by Berger and Brodsky
\cite{BergerBrodsky,BergerZPC4,Matsuda,BBKM,EHVV}.

The higher-twist charmonium production mechanism
which we present here is closely related to
the higher-twist Drell-Yan mechanism. We have chosen to
study the $C=+1$ states because they couple to two gluons, which makes this
the simplest case to calculate.
We shall show that
the $\chi_{1,2}$ are produced longitudinally polarized at large \xf. For the
$\chi_2$, this is strikingly different from the usual leading-twist
gluon-fusion mechanism, $gg \into \chi_2$, which only produces transversely
polarized $\chi_2$ \cite{VHBT,Ioffe}. For the $\chi_1$, on the other hand, the
leading- and higher-twist contributions have similar polarization properties,
so
that the signature of the higher-twist mechanisms is not as clear as for
the $\chi_2$.

Thus a measurement of the polarization of the $\chi_2$ will
provide important information about the production dynamics of charmonium.
An observation of the predicted
longitudinal polarization of the $\chi_{1,2}$ at
large \xfs would support the helicity conservation rule
between the pion and the
large \xfs system \cite{BergerBrodsky}, which
is already supported by data
for the \jps and the Drell-Yan virtual photons.

In Section 2 of this article we outline the calculation of the charmonium
production cross sections. We illustrate their \xfs dependence
in $\eta_c$ production, in which case compact analytical expressions can be
derived. In Section 3, we present numerical results for the polarization
of the $\chi_{1,2}$. In Section 4, we summarize our results
and discuss the relation of this model to other models of
charmonium production.

\section{Outline of the calculation}

The production of systems that carry a large momentum fraction \xfs requires a
perturbative momentum exchange involving all the valence partons of the
projectile \cite{BergerBrodsky,Gunion}. Thus a
correlation scale is present in the reaction. The cross sections
become suppressed by powers of the ratio
of this correlation scale and the hard momentum scale, \eg $f_\pi^2/m_c^2$ in
charm production from pions. These mechanisms are therefore of higher twist
and are usually neglected. At large \xf, however, inverse powers of
$(1-x_F)$ may compensate the suppression, and higher-twist mechanisms may
become dominant \cite{BHMT}.

Here, we present a higher-twist mechanism for $C=+1$ charmonium
production.
Our mechanism is described by the Feynman diagram in Fig. 1.
Note that whereas the
mechanism of Ref. \cite{BergerBrodsky}
produces both the
higher-twist component of the cross section
and the perturbative tail of
the leading-twist component,
our mechanism
is purely higher twist.
This is evident from the Feynman diagrams.
As in the Drell-Yan
case, we treat the interaction with the nuclear target simply as the
annihilation of one of the pion's valence quarks with an on-shell quark from
the target. This is the simplest higher-twist mechanism for charmonium
production, but not necessarily the most important one.
There may be a significant contribution from higher-order processes
where a gluon from the target interacts with the pion,
because at high energies and large \xfs the parton
distribution of the target is sampled at small $x_{\rm target}$.
Moreover, the simplifying assumption that the
target just contributes one parton implies that the
cross section is factorizable in terms of the target parton
distribution.
An analysis
\cite{HVS} of \jps production data \cite{Badier} suggests that there are
important non-factorizable production mechanisms.

We treat the pion bound-state effects as in Ref. \cite{LepageBrodsky} and the
charmonium binding as in the colour-singlet model \cite{KuhnKaplanSafiani}.
In other words,
the amplitude for the reaction $\pi^- + u \into d + \state$ is written as a
convolution of three factors: the pion distribution amplitude
$\phi(z)$, the perturbative-QCD amplitude for $ \ubar
d + u \into d + \ccbar$, and the non-relativistic
wavefunction $\Phi(P,q)$ of the
charmonium state. Here, $z$ is the light-cone momentum fraction carried by the
$\ubar$ quark in the pion, $P$ is the total four-momentum of the charmonium
state, and $2q$ is the relative
four-momentum of the charm quarks. The amplitude is
\beqa
  A(\pi^- + u \into d + ^{2S+1}L_J)
  & = & \frac{4\pi\als\sqrt{2}f_\pi}{9}
        \; \ubar_C(p_d) \gamma^{\mu_2} \gamma_5 \sla{p}_\pi
        \gamma^{\mu_1} u_C(p_u) \nonumber \\
  &   & \times \int dz \; \frac{\phi(z)}{zs (1-z)u} \;
        A_{\mu_1 \mu_2} (gg \into \state),
\eeqa
where $s =
(\ppi + p_u)^2$, $u = (\ppi - p_d)^2$, $C$ is a colour index, and
$A_{\mu_1 \mu_2}$ is the truncated amplitude for $gg \into \state$, which is
the convolution of the  quarkonium wavefunction $\Phi(P,q)$ with the
hard amplitude ${\cal O}_{\mu_1 \mu_2}$ for $gg \into \ccbar$,
\beq
  {\cal O}_{\mu_1 \mu_2} =
  \frac{2\pi\als}{\sqrt{3}} \left[ \gamma_{\mu_1} \;
  \frac{\frac{1}{2}\sla{P}-\sla{k}_1+\sla{q}+m_c}{(\frac{1}{2}P-k_1+q)^2 -
  m_c^2} \; \gamma_{\mu_2} + (1 \lra 2) \right],
\eeq
as explained in Ref. \cite{KuhnKaplanSafiani} (an example is given in
eq. (\ref{etacamplitude}) below).
The convolutions of the pion distribution amplitude with
${\cal O}_{\mu_1 \mu_2}$ and its derivative with respect to
$q$, evaluated at the non-relativistic limit
$q=0$, are
\beqa
  \left[ \int dz \frac{\phi(z)}{z(1-z)} {\cal O}_{\mu_1 \mu_2} \right]_{q=0}
  & = & \frac{4\pi\als}{\sqrt{3}(s-u)} \left[ I_0(z_0)
        G^{(0+)}_{\mu_1 \mu_2} + I_1(z_0) G^{(1-)}_{\mu_1 \mu_2} \right], \\
  \left[ \frac{\partial}{\partial q^\alpha} \int dz \frac{\phi(z)}{z(1-z)}
  {\cal O}_{\mu_1 \mu_2} \right]_{q=0}
  & = & \frac{8\pi\als}{\sqrt{3}(s-u)^2} \left\{
        2p_{\pi\alpha} \left[ I_0(z_0) G^{(0-)}_{\mu_1 \mu_2} - I_1(z_0)
        G^{(1+)}_{\mu_1 \mu_2} \right] \right.
        \nonumber \\
  &   & \mbox{} + (s-u) I_0(z_0) \left[ g_{\alpha\mu_1} \gamma_{\mu_2}
        + g_{\alpha\mu_2} \gamma_{\mu_1} \right]
        \nonumber \\
  &   & \left. \mbox{}+ 2 (p_{u\alpha} + z_0 p_{\pi\alpha})
        \left[ I_0'(z_0) G^{(0-)}_{\mu_1 \mu_2}
        - I_1'(z_0) G^{(1+)}_{\mu_1 \mu_2} \right] \right\},
        \nonumber \\
  &   &
\eeqa
where
\beqa
  I_n(z_0)                 & = & \int dz \frac{\phi(z)}{z(1-z)}
                                 \frac{z^n}{z-z_0+i\epsilon}, \\
  z_0                      & = & \frac{M^2-u}{s-u}, \\
  G^{(0\pm)}_{\mu_1 \mu_2} & = &  P_{\mu_1} \gamma_{\mu_2}
                                  - \gamma_{\mu_1} \sla{p}_u \gamma_{\mu_2}
                                  \pm [P_{\mu_2} \gamma_{\mu_1} -
                                  \gamma_{\mu_2} (\sla{p}_\pi-\sla{p}_d)
                                  \gamma_{\mu_1}], \\
  G^{(1\pm)}_{\mu_1 \mu_2} & = &  \gamma_{\mu_2} \sla{p}_\pi \gamma_{\mu_1}
                                  \pm \gamma_{\mu_1} \sla{p}_\pi
                                  \gamma_{\mu_2}.
\eeqa The singularities at $z=0,1$ from the gluon
propagators are expected to be cancelled by the distribution amplitude
$\phi(z)$, whereas the singularity at $z=z_0$ from the heavy quark propagator
is regularized by the $+i\epsilon$ prescription, which gives the amplitude a
non-trivial phase \cite{Kronfeld}. Below, we
shall present results for the symmetric (asymptotic) \cite{LepageBrodsky}
and two-humped \cite{Chernyak} distribution amplitudes\footnote{The
simplest model for the pion's valence state,
$\phi(z) = \delta(z-1/2)$, leads to rather non-realistic shapes of the cross
sections since the $+i\epsilon$ prescription fails to regularize the
singularity of the quark propagator
if the distribution amplitude does not vary
smoothly over the pole. The same is actually true of the higher-twist
contribution \cite{BergerBrodsky} to the Drell-Yan cross section, although the
singularity is cancelled when one considers angular distributions.},
\beqa
  \phi(z) & = & 6z(1-z), \\
  \phi(z) & = & z(1-z) [ 26 - 100 z(1-z) ].
\eeqa

We shall first study
$\eta_c$ production with a symmetric pion
distribution amplitude, in which case
the analytic expressions are quite compact.
For the $\eta_c$, the relation between $A_{\mu_1 \mu_2}$ and  ${\cal O}_{\mu_1
\mu_2}$ is \cite{KuhnKaplanSafiani}
\beq
  A_{\mu_1 \mu_2} = \frac{R_S(0)}{\sqrt{16\pi M}}
                    {\rm Tr} [ {\cal O}_{\mu_1 \mu_2} \gamma_5  (\sla{P}-M) ],
                    \label{etacamplitude}
\eeq
where $M = 2m_c$ is
the mass of the charmonium state, and $R_S(0)$ is the
value of the S-wave wavefunction at the spatial origin. The square of the $\pi
u \into d \eta_c$ amplitude, averaged over the spin and colour of the $u$
quark, is
\beqa
  \overline {|A|^2} & = & \frac{32(4\pi\als)^4 |R_S|^2 f_\pi^2}{27\pi M}
                          \frac{1}{su(s-u)^2} \left\{ -s^2 (L^2 + \pi^2)
\right.
                          \nonumber \\
                    &   & \left. \mbox{}
                          + 2s(s+u) \left[ L(1+z_0 L) + \pi^2 z_0 \right]
                          - (s-u)^2 \left[ (1+z_0 L)^2 + \pi^2 z_0^2 \right]
                          \right\},
\eeqa
where $L \equiv \ln [(s-M^2)/(M^2-u)]$. The differential cross section is
\beq
  \frac{d\sigma}{dx d\ptr^2} = \frac{1}{16\pi\stot}
                                  \left[ f_{u/N}(s/\stot )
                                  + f_{\dbar/N}(s/\stot ) \right]
                                  \frac{\overline {|A|^2}}{x(1-x)s},
\eeq
where $\stot  = (\ppi + p_{\rm nucleon})^2$. In this and the
following formulae, we use the light-cone momentum fraction
$x = (P^0+P^3)/(p_\pi^0+p_\pi^3)$ instead of the longitudinal momentum fraction
\xf. There is a simple relation between these two:
$x_F = x - (M^2+\ptr^2)/(x\stot)$. The two contributions are from
$\pi + u \into d + \state$ and $\pi + \dbar \into \ubar + \state$,
respectively.

In the limit $M^2 \gg \ptr^2$ with $x$ finite, the cross section
simplifies to
\beqa
  \frac{d\sigma}{dx d\ptr^2}
  & = & \frac{512\pi^2 \als^4 |R_S|^2 f_\pi^2}{27 M^5 \stot}
        \left[ f_{u/N} \left( \frac{M^2}{x\stot} \right)
        + f_{\dbar/N} \left( \frac{M^2}{x\stot} \right) \right]
        \frac{1}{\ptr^2}
        \nonumber \\
  &   & \times x \left\{ \left[ 1 - (1-x)\ln\frac{1-x}{x} \right] ^2 +
        \pi^2 (1-x)^2 \right\}. \label{largeMlimit}
\eeqa
Now, taking the limit $x \into 1$ in (\ref{largeMlimit}) means taking the
double limit
\beq
  M^2 \gg \ptr^2, \;\;\; x \into 1
  \;\; {\rm with} \;\; M^2(1-x) \gg \ptr^2,
  \label{BBlimit}
\eeq
\ie the same limit as for the Drell-Yan process in Ref. \cite{BergerBrodsky}.
Although the limit
(\ref{BBlimit}) is not quite reached in present-day pion-nucleon reactions
\cite{EHVV}, it
illustrates the behaviour of the
$\ptr$-integrated cross section at large $x$ since it takes into account the
fact that as $x$ is increased, the dominant region of the integration shifts to
smaller \spt. The cross section (\ref{largeMlimit}) simplifies to
\beq
  \frac{d\sigma}{dx d\ptr^2} =
    \frac{512\pi^2 \als^4 |R_S|^2 f_\pi^2}{27 M^5 \stot}
    \left[ f_{u/N} \left( \frac{M^2}{\stot} \right)
    + f_{\dbar/N} \left( \frac{M^2}{\stot} \right) \right]
    \frac{1}{\ptr^2}.
    \label{atBBlimit}
\eeq
The integral $\int d\ptr^2$ over the perturbative region where
$\ptr^2/(1-x) \gg \Lambda_{\rm QCD}^2$ brings a logarithmic factor.
Apart from this logarithm, the
higher-twist contribution is suppressed by $f_\pi^2/M^2$ with
respect to the leading-twist,
leading-order contribution from the subprocess $gg \into \eta_c$
\cite{BaierZPC19}, which is
\beq
  \int_0^\infty d\ptr^2 \frac{d\sigma_{\rm LT}}{dx d\ptr^2}
     = \frac{\pi^2\als^2 |R_S|^2}{3M^3 \stot} \; \frac{1}{x} \;
       f_{g/\pi}(x) f_{g/N}(\frac{M^2}{x\stot }).
\eeq
{}From (\ref{atBBlimit}) we also see that
$d\sigma/dx$ remains constant as $x \into 1$, up to the logarithmic factor from
the $\ptr$ integration. The leading-twist contribution, on the other hand,
falls off according to the power behaviour of the pion's gluon distribution,
\ie $(1-x)^3$ from the spectator counting rules\footnote{Experimental
determinations of the large $x$ behaviour of the gluon distribution
\cite{OwensPRD30,Aurenche,Sutton} are based on fitting a purely
leading-twist model to Drell-Yan, prompt-photon and/or charmonium data.
The resulting powers of $(1-x)$ therefore partly reflect a hardening of
the cross section due to the higher-twist component.}
\cite{Sivers}. At very
large $x$, this will compensate the suppression by $f_\pi^2/M^2$, and the
higher-twist component will become dominant.

\section{The polarization of charmonium}

We now consider the production of the $\chi_{1,2}$ states
within our higher-twist model. The cross sections
for
absolute values of the helicity
$|\lambda| = 0,1,2$ can be extracted by using the
covariant polarization sums given in \eg
Ref. \cite{Cho}. However, the analytical expressions
that are obtained by using a symbolic
manipulation program such as Reduce \cite{Reduce} are too lengthy
to be given here (some of them contain over a thousand terms)
or even to be used in numerical integration routines.
Instead, we have evaluated the double-differential polarized cross sections
$d\sigma/dxd\ptr^2 \; (|\lambda|=0,1,2)$ numerically at a characteristic
value of the transverse
momentum, $\ptr^2 = M^2(1-x)$.

In Fig. 2 we plot the ratio
\beq
  R(1) = \frac{\sum_{\lambda=\pm 1}d\sigma/dxd\ptr^2 \; (\lambda)
         }{d\sigma/dxd\ptr^2 \; (\lambda=0)}
  \label{R1}
\eeq
of the
polarized cross sections for the $\chi_1$ as a function of \xf, using the
symmetric and two-humped distribution amplitudes. The ratio of the
polarized leading-twist cross sections $d\sigma/dx$ \cite{VHBT} is also shown.
In Fig. 3 we plot the ratios
$R(1)$ and $R(2)$
for the $\chi_2$. The leading-twist, leading-order
gluon-fusion mechanism $gg \into \chi_2$ gives $d\sigma/dx \; (0,\pm 1) = 0$,
up to corrections of the order of 15 \% from transverse
momentum smearing of the pion \cite{VHBT}.

As $x_F \into 1$, the contribution from our mechanism to both the $\chi_1$
and $\chi_2$ cross sections becomes longitudinally polarized ($\lambda=0$).
For the $\chi_2$, this is in striking
contrast to the leading-twist prediction. For the $\chi_1$, however,
the polarization of the higher-twist component is
qualitatively similar to
that of the leading-twist component,
so that polarization cannot be used to discriminate between leading- and
higher-twist mechanisms as for the $\chi_2$.

\section{Discussion}

When a charmonium state is produced at a very large momentum fraction \xf,
all of the projectile hadron's momentum must be transferred to the
charm quarks. This requires a higher-twist mechanism, where all
the valence quarks of the projectile couple perturbatively to the charm quarks.
The higher-twist nature of these processes is due to
the soft correlation scale, \eg $f_\pi$,
from the integration over the transverse momentum distribution of the valence
state. In this article, we have presented a higher-twist mechanism for the
production of the $\eta_c$ and $\chi_J$ charmonia in pion-nucleon collisions.
The contribution from this mechanism
dominates
over the leading-twist contribution at large \xf.
As seen from eq. (\ref{atBBlimit}) for the $\eta_c$,
the higher-twist cross section is almost flat
as $x \into 1$, whereas the leading-twist cross section falls off as
$(1-x)^3$.

In our model, both the $\chi_1$ and $\chi_2$
are produced longitudinally
polarized at large \xf.
For the $\chi_2$, this result
is in striking contrast to the leading-twist prediction of
purely transverse polarization; for the $\chi_1$, the polarization of the
leading- and higher-twist contributions is
qualitatively similar.
A direct measurement of the polarization of the $\chi_2$ would
give important information about the production mechanisms of
charmonium\footnote{The decay angular distributions of P-wave charmonia
have actually been
measured recently \cite{AntoniazziPRD49}. However, no separation was made
between the $\chi_1$ and $\chi_2$ in the angular distribution analysis, and due
to limited statistics, the errors are still
quite large. With the present statistics, it is also impossible to determine
the polarization as a function of the momentum fraction \xf.}.

The mechanism considered in this paper is the simplest but not necessarily the
most important of many possible higher-twist production mechanisms.
In mechanisms
such as this and the Drell-Yan mechanism of Ref. \cite{BergerBrodsky}, where
the valence state of the pion interacts with a single parton from the target
to produce a large \xfs system in the final state,
the parton distribution of the target is
at high energies
sampled at small values of $x_{\rm target}$. Due to the dominance of gluon
distributions over quark distributions at small $x_{\rm target}$,
there may be significant contributions from higher-order (in \as)
mechanisms where the target contributes a gluon instead of a quark.

Mechanisms where the target contributes a single parton are factorizable in
terms of the parton distribution of the target. Thus they would not bring about
the violation of factorization seen in the reaction $\pi + A \into J/\psi + X$
\cite{HVS}. Instead, the nuclear number dependence of the cross section would
be consistent with shadowing, as
is experimentally the case
in the Drell-Yan reaction. More general
higher-twist mechanisms outlined in Ref. \cite{BHMT}, where a
spectator valence
quark interacts softly with the target, would give factorization-violating
cross sections.

In this article, we have compared the leading- and higher-twist components
of the charmonium production cross sections within the colour-singlet model.
There is also a potentially important contribution
to the production
of the $\chi_J$ states from mechanisms which involve non-perturbative
transitions between colour-singlet and colour-octet $\QQbar$ states
\cite{BodwinBraatenLepagePRD51}.
These mechanisms are necessary in a rigorous
calculation of the $\chi_J$ decay widths \cite{BodwinBraatenLepagePRD46},
which in the colour-singlet model are
infrared divergent at $\order(\als^3)$. In
the rigorous analysis, there is a contribution
proportional to $\als^2 < \chi_J | \order_8(^3 S_1) | \chi_J >$, where the
infrared singularity has been absorbed into the non-perturbative
matrix element $< \chi_J | \order_8(^3 S_1) | \chi_J >$.
In $\chi_J$ production, there is a corresponding contribution from
the non-perturbative production matrix element
$ < 0 | \order_8^{\chi_J} (^3 S_1) | 0 > $. The magnitude of this matrix
element can be measured in $B$-meson decays into $\chi_J$ charmonium
\cite{BodwinBraatenYuanLepage,BraatenFleming}.
In our mechanism, where the $\chi_J$ states couple to two gluons, there is no
infrared singularity that would signal the breakdown of the colour-singlet
model. Nevertheless, the octet contribution is of the same order in the
perturbative and non-relativistic expansions as the singlet contribution. To
estimate the importance of colour-octet mechanisms in $\chi_J$ production at
higher twist, one should calculate the cross section for the process
$\pi^- + u \into \; ^3 S_1 + d$ where the $\QQbar(^3 S_1)$ heavy
quark-antiquark
pair is in a colour-octet state. We postpone the analysis of these
contributions to future work.

In spite of these reservations, the present calculation
illustrates
the perturbative dynamics behind the hadron helicity conservation at large
\xf, which has been observed in the Drell-Yan process \cite{Guanziroli,Conway}
and in \jps production \cite{Biino}.

Finally, we note that the present mechanism and other
similar mechanisms where the
charmonium state couples to virtual gluons allow the production of $\chi_1$
at the same order in \as\ as $\chi_2$. In the leading
twist case, the production of $\chi_1$ in the leading-order subprocess of
on-shell gluon fusion, $gg \into \chi_1$, is forbidden by
the Landau-Yang theorem
\cite{Yang}, and consequently the ratio of the total $\chi_1$ and $\chi_2$
production cross sections is predicted to be small. The
experimental data shows equal cross sections \cite{AntoniazziPRL70}. If
higher-twist production mechanisms
are important
down to low \xf, as the
failure of the leading-twist predictions of charmonium polarization \cite{VHBT}
seems to indicate, they could explain the total cross
section anomaly.

\bigskip

This work was supported by the Academy of Finland under project number 8579.
The author wishes to thank NORDITA and the theory group of SLAC for their warm
hospitality during his visits in December 1994 and January 1995, when much of
the work reported in this article was done. He is grateful to P. Hoyer, S. J.
Brodsky and W.-K. Tang for important discussions.

\pagebreak

\bigskip \centerline{\large FIGURE CAPTIONS} \vspace{1cm}

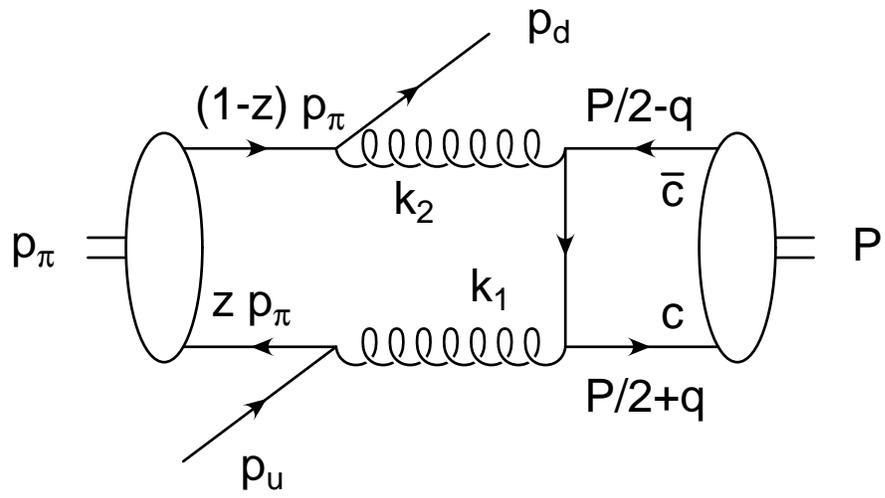
{\bf Figure 1.}
One of the two Feynman diagrams that describe
the higher-twist charmonium production mechanism
considered in this article. The four-momenta flow from left to right.
The other diagram is obtained by crossing the gluon lines.

\bigskip

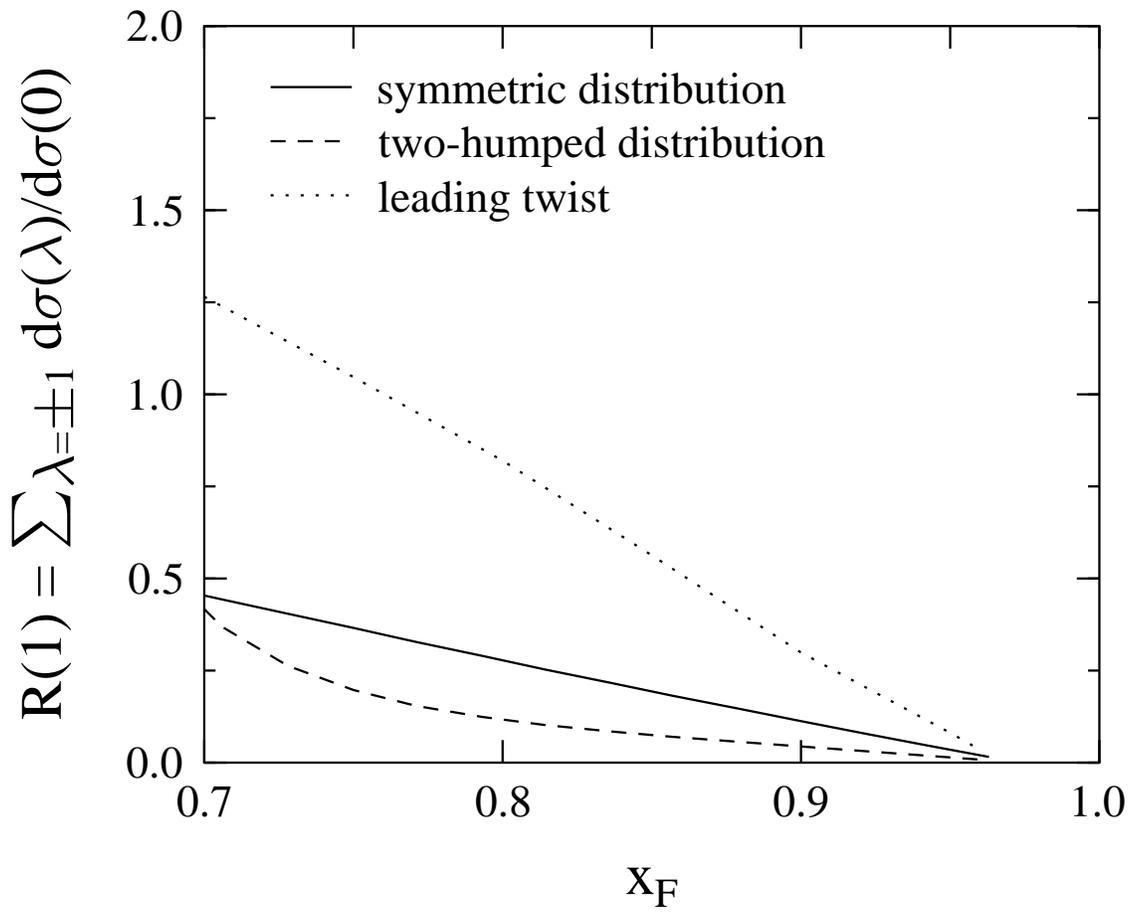
{\bf Figure 2.}
The ratio $R(1) = \sum_{\lambda=\pm 1} d\sigma(\lambda)/d\sigma(0)$
of the polarized $\chi_1$ production
cross sections plotted as a function of \xf. The higher-twist contributions
have been evaluated at a characteristic
value of the transverse momentum, $\ptr^2 = M^2(1-x)$,
using the symmetric and two-humped pion distribution amplitudes.
The leading-twist
contribution \cite{VHBT} has been integrated over \spt.

\bigskip

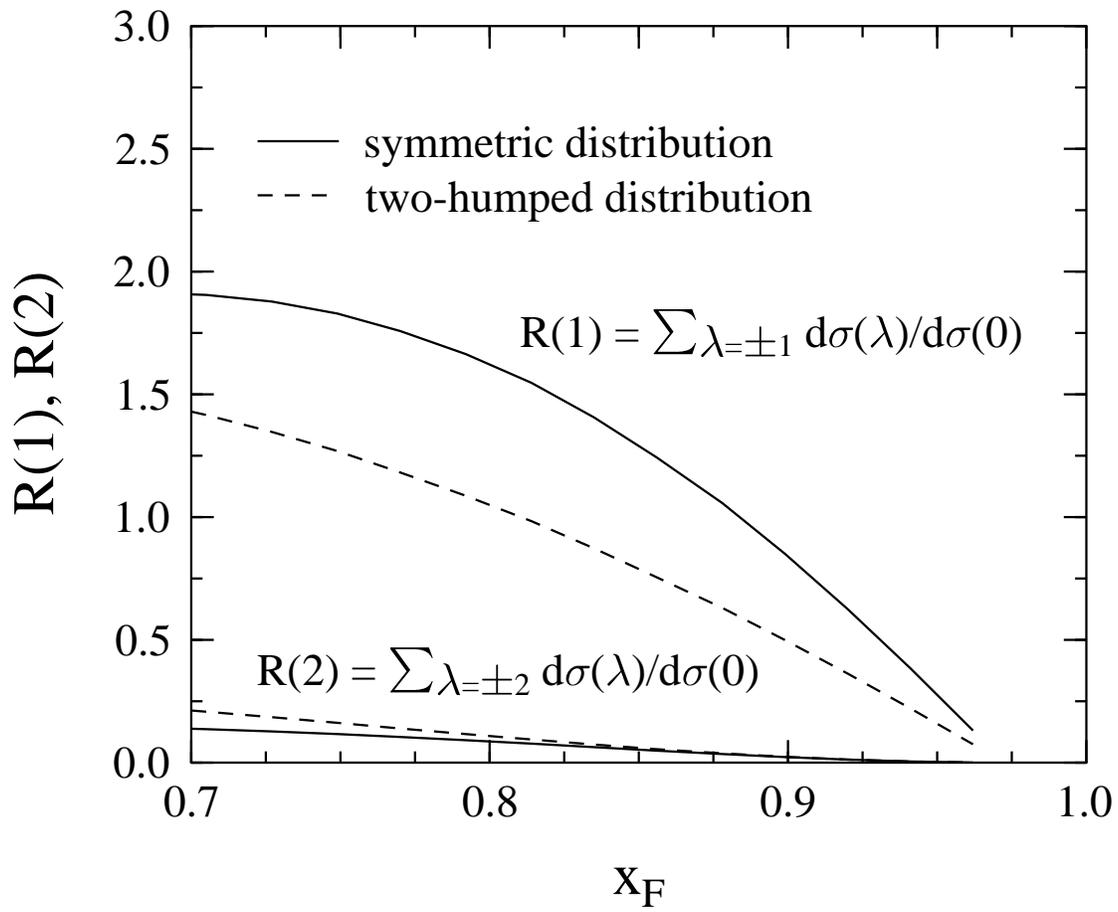
{\bf Figure 3.}
The ratios  $R(1) = \sum_{\lambda=\pm 1} d\sigma(\lambda)/d\sigma(0)$
and  $R(2) = \sum_{\lambda=\pm 2} d\sigma(\lambda)/d\sigma(0)$
of the polarized $\chi_2$ production
cross sections, plotted as a function of \xf.
The higher-twist contributions, which are shown here,
have been evaluated at a characteristic
value of the transverse momentum, $\ptr^2 = M^2(1-x)$,
using the symmetric and two-humped pion distribution amplitudes.
The leading-twist prediction, with transverse momentum
smearing and higher-order corrections neglected, is $d\sigma(0,\pm 1)=0$.

\end{document}